\documentclass[final,3p,times,12pt]{elsarticle}

\usepackage{graphicx}
\usepackage{amssymb}
\usepackage{amsmath}
\usepackage{booktabs}

\biboptions{comma,square,sort&compress}

\journal{Physics Letters B}

\begin{document}
\begin{frontmatter}

\vbox{\hbox{ADP-12-35/T802}}

\title{Strong contribution to octet baryon mass splittings}

\author{P.E.~Shanahan}

\author[]{A.W.~Thomas\corref{cor1}}

\author{R.D.~Young}

\address{ARC Centre of Excellence in Particle Physics at the Terascale and CSSM, School of Chemistry and Physics, University of Adelaide,
  Adelaide SA 5005, Australia}

\cortext[cor1]{Corresponding author}

\begin{abstract}

We calculate the $m_d-m_u$ contribution to the mass splittings in baryonic isospin multiplets using SU(3) chiral perturbation theory and lattice QCD. Fitting isospin-averaged perturbation theory functions to PACS-CS and QCDSF-UKQCD Collaboration lattice simulations of octet baryon masses, and using the physical light quark mass ratio $m_u/m_d$ as input, allows $M_n-M_p$, $M_{\Sigma^-}-M_{\Sigma^+}$ and $M_{\Xi^-}-M_{\Xi^0}$ to be evaluated from the full SU(3) theory. The resulting values for each mass splitting are consistent with the experimental values after allowing for electromagnetic corrections. In the case of the nucleon, we find $M_n-M_p= 2.9 \pm 0.4 \textrm{ MeV}$, with the dominant uncertainty arising from the error in $m_u/m_d$.

\end{abstract}


\end{frontmatter}

\section{Introduction}
\label{}

The physical mass splittings between members of baryonic isospin multiplets have been measured extremely precisely~\cite{Mohr2012,pdg}:
\begin{subequations}
\label{eq:exptsplit}
\begin{alignat}{2}
&M_n - M_p & \;= & \; 1.2933322(4)~\textrm{MeV},\\
&M_{\Sigma^-}-M_{\Sigma^+} & = & \; 8.079(76)~\textrm{MeV}, \\
&M_{\Xi^-}-M_{\Xi^0} & = & \; 6.85(21)~\textrm{MeV}.
\end{alignat}
\end{subequations}
However, the decomposition of each into its two components, arising from electromagnetic effects and the $d-u$ quark mass difference, is less well known.
Clearly, once one contribution has been well determined the other can be inferred from the total.

In recent years, several research groups have presented lattice determinations of both the QCD contribution to the baryon mass splittings, e.g.,~\cite{Beane2006,Blum2010,WalkerLoud2010,Divitiis2011,Horsley2012}, and the electromagnetic contribution, e.g.,~\cite{Duncan1996,Basak2008,Portelli2011,Glaessle2011}.
This work uses SU(3) chiral perturbation theory to determine estimates of the strong contribution, by fitting currently available isospin-averaged lattice calculations~\cite{Aoki2009,Bietenholz2011} of the octet baryon masses. We do not consider electromagnetic effects.

Our evaluation of the strong contribution to $M_n-M_p$ is of particular interest in the light of recent results which suggest that the accepted value for the electromagnetic contribution, namely $\Delta_{\textrm{EM}} =  0.76 \pm 0.30$~MeV~\cite{Gasser1982}, may be too small. Walker-Loud~\textit{et al.} (WLCM) claim to find an omission in the traditional analysis and present a larger value of $1.30 \pm 0.03 \pm 0.47$~MeV~\cite{Walker-Loud2012}. From these estimates, one infers strong isospin breaking contributions of $\Delta_{m_d-m_u} = 2.05 \pm 0.30$~MeV (traditional) and $2.60 \pm 0.47$~MeV (WLCM) respectively. Clearly, independent theoretical estimates of the size of the strong contribution to $M_n-M_p$, such as reported here, are of considerable value.

\section{Method}

\subsection{Isospin-averaged fit}

The fit to isospin-averaged PACS-CS lattice results~\cite{Aoki2009} which we use for this work has been reported on in previous papers; we refer to~\cite{Young2010,Shanahan2011,Shanahan2012} for details.
In brief, we use a standard heavy-baryon chiral perturbation theory formulation with a finite-range regularization scheme (FRR), discussed in~\cite{Stuckey1997,Donoghue1999,Young2002,YOung2003,Leinweber2004}. The mass of a baryon $B$ in the chiral expansion is written as
\begin{equation}
\label{massequation}
M_B=M^{(0)}+ \delta M_B^{(1)}+ \delta M_B^{(3/2)}+ \ldots,
\end{equation}
where the leading term $M^{(0)}$ denotes the degenerate mass of the octet baryons in the SU(3) chiral limit and is independent of the quark mass matrix $\mathcal{M}_q$ and baryon $B$. The notation $M_B^{(n)}$ denotes the contribution to the octet baryon mass at order $\mathcal{M}_q^{(n)}$.

The correction linear in the quark masses can be expressed as
\begin{equation}
\delta M_B^{(1)} = -C_{Bl}^{(1)} B m_l - C_{Bs}^{(1)} B m_s,
\label{eq:linear}
\end{equation}
with the coefficients given in Table~\ref{tab:linear}~\cite{WalkerLoud2004}.
The leading loop corrections which contribute to $\delta M_B^{(3/2)}$ are made explicit in~\cite{Shanahan2011} and include both octet and decuplet baryon intermediate states.
These contributions take the form:
\begin{equation}
\delta M_B^{(3/2)} = -\frac{1}{16\pi f^2} \sum_\phi \left[\chi_{B\phi}I_R(m_\phi,0,\Lambda)+\chi_{T\phi}I_R(m_\phi,\delta,\Lambda)\right] \, ,
\label{eq:3}
\end{equation}
where the meson loops involve the integrals:
\begin{equation}
I_R=\frac{2}{\pi}\int dk \frac{k^4}{\sqrt{k^2+m_{\phi}^2}(\delta + \sqrt{k^2+m_{\phi}^2})}u^2(k)-b_0-b_2m_{\phi}^2.
\label{eq:4}
\end{equation}
The subtraction constants, $b_{0,2}$, are defined so that the parameters $M^{(0)}, C_{Bl}^{(1)}$ and 
$C_{Bs}^{(1)}$ are renormalized (explicit expressions may be found in Ref.~\cite{Leinweber2004}, or can be readily evaluated numerically by Taylor expanding the integrand in $m_{\phi}^2$).

\begin{table}
\begin{center}
\begin{tabular}{c c c}
\toprule
$B$ & $C_{Bl}^{(1)}$ & $C_{Bs}^{(1)}$ \\ \midrule
$N$ & $2\alpha+2\beta+4\sigma$ & $2\sigma$ \\
$\Lambda$ & $\alpha + 2\beta + 4\sigma$ & $\alpha+2\sigma$ \\
$\Sigma$ & $\frac{5}{3}\alpha + \frac{2}{3}\beta + 4\sigma$ & $\frac{1}{3}\alpha+ \frac{4}{3}\beta + 2\sigma$ \\
$\Xi$ & $\frac{1}{3}\alpha+\frac{4}{3}\beta+4\sigma$ & $\frac{5}{3}\alpha+\frac{2}{3}\beta+2\sigma$ \\ \bottomrule
\end{tabular}
\caption{Coefficients of terms linear in the non-strange quark mass, 
$Bm_l \rightarrow m_\pi^2/2$, and the strange quark mass, $Bm_s \rightarrow (m_K^2 - m_\pi^2/2)$,
expressed in terms of the leading quark-mass insertion parameters $\alpha$, $\beta$ and $\sigma$.}
\label{tab:linear}
\end{center}
\end{table}

We retain the octet-decuplet mass splitting $\delta$ in numerical evaluations, setting $\delta=0.292$~GeV to the physical $N-\Delta$ splitting. The baryon-baryon-meson coupling constants are taken from phenomenology; $D+F=g_A=1.27$, $F=\frac{2}{3}D$ and $\mathcal{C}=-2D$, and $f$ is set to $f=0.0871$~GeV, a chiral perturbation theory estimate for the pion decay constant in the SU(3) chiral limit~\cite{Amoros2001}. The fit parameters are the octet baryon mass in the chiral limit $M^{(0)}$, the SU(3) chiral symmetry breaking parameters $\alpha$, $\beta$, $\sigma$, and the finite-range regulator mass $\Lambda$, and correspond to those in Refs.~\cite{Shanahan2011,Shanahan2012}.

The fit to the PACS-CS baryon octet data is shown in~\cite{Shanahan2012}, and a comparison between the experimental values and the octet baryon masses evaluated at the physical point is given in Table~\ref{tab:barmasses}.


\begin{table}[hftb]
\begin{center}
\begin{tabular}{c r@{.}l c}
\toprule
$B$ & \multicolumn{2}{c}{Mass (GeV)} & Experimental \\ 
\midrule
$N$ & 0 &959(24)(9) & 0.939  \\
$\Lambda$ & 1 &129(15)(6) & 1.116  \\
$\Sigma$ & 1 &188(11)(6) & 1.193  \\
$\Xi$ & 1 &325(6)(2) & 1.318 \\
\bottomrule
\end{tabular}
\caption{Extracted masses for the octet baryons. The first uncertainty quoted is statistical, while the
  second allows for variation of the form of the UV regulator and 10\% deviation of $f$, $F$, $\mathcal{C}$, and $\delta$ from their central values. The
  experimental baryon masses are shown for comparison.}
\label{tab:barmasses}
\end{center}
\end{table}

\subsection{Evaluation of SU(2) mass splittings}

A feature of SU(3) chiral perturbation theory is that the same coefficients appear in the baryon mass expansion both when including and excluding SU(2) breaking effects. These coefficients, once evaluated by fitting to the isospin-averaged lattice results as done above, can thus be used to provide information about the SU(2) breaking mass splittings. That is, we can use $N_f=2+1$ flavor lattice simulations, which are currently available, and are computationally cheaper than the SU(2)-broken $N_f=1+1+1$, to derive our results.

To calculate the baryon mass splittings, we modify the SU(3) chiral perturbation theory expansions used in the previous section to allow for a  non-zero light quark mass splitting: $m_d-m_u \ne 0$. While this is a straightforward extension, we note that it generates a $\pi^0 \eta$ mixing term in the SU(3) Lagrangian. The fields must thus be diagonalized into the mass basis via a field rotation.

To be explicit, we recall the usual definition of the meson field:
\begin{equation}
\Sigma = \textrm{exp}\left(\frac{2 i \Phi}{f}\right)=\xi^2, \; \Phi = \frac{1}{\sqrt{2}}\left( \begin{array}{ccc}
\frac{1}{\sqrt{2}}\pi^0+\frac{1}{\sqrt{6}}\eta & \pi^+ & K^+ \\
\pi^- & -\frac{1}{\sqrt{2}}\pi^0+\frac{1}{\sqrt{6}}\eta & K^0 \\
K^- & \overline{K}^0 & -\frac{2}{\sqrt{6}}\eta \end{array} \right),
\end{equation}
and meson Lagrangian:
\begin{equation}
\label{mesonlagrangian}
\mathcal{L}_{\textrm{{eff}}}=\frac{f^2}{8}\textrm{Tr}(\partial^\mu \Sigma^\dagger \partial_\mu \Sigma) + \lambda \mathrm{Tr}(\mathcal{M}_q( \Sigma^\dagger + \Sigma)),
\end{equation}
where $\mathcal{M}_q = \textrm{diag}(m_u,m_d,m_s)$ is, as above, the quark mass matrix.

Expanding this Lagrangian in powers of the meson field, the mass term can be written as 
\begin{subequations}
\begin{align}
\mathcal{L}_{\textrm{kin}}= &B\textrm{Tr}(\mathcal{M}_q \Phi^2)\\
 = & B(m_u+m_d)(\pi^+ \pi^-) + B(m_s+m_d)(K^0 \overline{K}^0) \\ \nonumber
 & + B(m_s+m_u)(K^+ K^-) + \frac{B}{2}(m_u+m_u)((\pi^0)^2) \\ \nonumber
& +\frac{B}{6}(m_d+m_u+4m_s)(\eta^2)+\frac{B}{\sqrt{3}}(m_u-m_d)(\eta \pi^0).
\end{align}
\end{subequations}
Clearly, for $m_u \ne m_d$, mixing occurs between the $\pi^0$ and $\eta$.

To identify the meson masses we remove this mixing and bring the kinetic term into the canonical form via a field rotation:
\begin{subequations}
\begin{align}
\pi^0 & \rightarrow \pi^0 \textrm{cos} \epsilon - \eta \textrm{sin} \epsilon,\\
\eta & \rightarrow \pi^0 \textrm{sin} \epsilon + \eta \textrm{cos} \epsilon,
\end{align}
\end{subequations}
where the mixing angle $\epsilon$ is given by
\begin{equation}
\textrm{tan} 2 \epsilon = \frac{\sqrt{3}~(m_d-m_u)}{2m_s-(m_d+m_u)}.
\end{equation}

After performing this rotation, the SU(3) meson masses take the form:
\begin{subequations}
\label{mesonmasses}
\begin{align}
m_{\pi^\pm}^2 & = B(m_u + m_d)\\
m_{\pi^0}^2 & =B(m_u+m_d)-\frac{2B}{3}(2m_s-(m_u+m_d))\frac{\textrm{sin}^2\epsilon}{\textrm{cos} 2 \epsilon}\\
m_{K^\pm}^2 & = B(m_s+m_u)\\
m_{K^0}^2 & = B(m_s+m_d)\\
m_{\eta}^2 & = \frac{B}{3}(4m_s+m_u+m_d)+\frac{2B}{3}(2m_s-(m_u+m_d))\frac{\textrm{sin}^2\epsilon}{\textrm{cos} 2 \epsilon},
\end{align}
\end{subequations}
where $m_{\pi^0}$ and $m_\eta$ now contain some dependence on the mixing angle $\epsilon$.

This extension generates a separate mass expansion, of the form of Equation~\ref{massequation}, for each member of the baryon octet. 
The terms linear in quark mass can be expressed as
\begin{equation}
\delta M_B^{(1)} = -C_{Bu}^{(1)} B m_u-C_{Bd}^{(1)} B m_d - C_{Bs}^{(1)} B m_s,
\label{eq:linearSplit}
\end{equation}
where the coefficients $C_{Bq}$ are given explicitly in Table~\ref{tab:linearsplit}. 

The loop contributions $\delta M_B^{(3/2)}$ have the same form as in the isospin-averaged case, with separate couplings and integrals for each of the mesons $\pi^\pm,\pi^0,K^\pm,K^0,\eta$. The $\pi^\pm$ and $K^\pm$ remain pairwise mass-degenerate. Of course, because of the redefinition of the meson fields, the baryon-baryon-meson couplings will also receive contributions depending on $\epsilon$, and are now complicated functions of quark mass and the coupling constants $F$, $D$, and $\mathcal{C}$. These are given in Tables~\ref{tab:deccoup} and~\ref{tab:octcoup}. As expected, setting $\epsilon \rightarrow 0$ returns the usual isospin-averaged functions.

\begin{table}[htbf]
\begin{center}
\begin{tabular}{c c c c}
\hline
$B$ & $C_{Bu}^{(1)}$ & $C_{Bd}^{(1)}$ &  $C_{Bs}^{(1)}$ \\ \hline
$p$ & $\frac{5}{3}\alpha+\frac{2}{3}\beta+2\sigma$ & $\frac{1}{3}\alpha + \frac{4}{3}\beta + 2\sigma$ &$2\sigma$ \\
$n$ &$\frac{1}{3}\alpha + \frac{4}{3}\beta + 2\sigma$  &$\frac{5}{3}\alpha+\frac{2}{3}\beta+2\sigma$ & $2\sigma$\\
$\Sigma^+$ & $\frac{5}{3}\alpha + \frac{2}{3}\beta + 2\sigma$ & $ 2\sigma$ &$\frac{1}{3}\alpha+ \frac{4}{3}\beta + 2\sigma$ \\
$\Sigma^-$ & $ 2\sigma$ & $\frac{5}{3}\alpha + \frac{2}{3}\beta + 2\sigma$ &$\frac{1}{3}\alpha+ \frac{4}{3}\beta + 2\sigma$ \\
$\Xi^0$ & $\frac{1}{3}\alpha+\frac{4}{3}\beta+2\sigma$ & $2\sigma$ & $\frac{5}{3}\alpha+\frac{2}{3}\beta+2\sigma$ \\
$\Xi^-$ & $2\sigma$ & $\frac{1}{3}\alpha+\frac{4}{3}\beta+2\sigma$  & $\frac{5}{3}\alpha+\frac{2}{3}\beta+2\sigma$ \\ \hline
\end{tabular}
\end{center}
\caption{Values for the terms linear in the up, down and strange quark masses, expressed in terms of the SU(3) breaking parameters $\alpha$, $\beta$ and $\sigma$.}
\label{tab:linearsplit}
\end{table} 

\begin{table}[htbf]
\begin{center}
\begin{tabular}{c c c c c c}
\hline
& \multicolumn{5}{c}{$\chi_{T\phi}\mathcal{C}^{-2}$} \\
 & $\pi^0$ & $\pi^\pm$ & $K^0$ & $K^\pm$  &$\eta$ \\ \hline
$p$ & $\frac{4}{9}\textrm{cos}^2\epsilon$ & $\frac{8}{9}$ & $\frac{2}{9}$ & $\frac{1}{9}$ & $\frac{4}{9}\textrm{sin}^2\epsilon$ \\
$n$ & $\frac{4}{9}\textrm{cos}^2\epsilon$ & $\frac{8}{9}$ & $\frac{1}{9}$ & $\frac{2}{9}$ & $\frac{4}{9}\textrm{sin}^2\epsilon$ \\
$\Sigma^+$ & $\frac{1}{9}(\textrm{cos}\epsilon + \sqrt{3}\textrm{sin}\epsilon)^2$ & $\frac{1}{9}$ & $\frac{2}{9}$ & $\frac{8}{9}$ & $\frac{1}{9}(-\sqrt{3}\textrm{cos}\epsilon + \textrm{sin}\epsilon)^2$ \\
$\Sigma^-$ & $\frac{1}{9}(-\textrm{cos}\epsilon + \sqrt{3}\textrm{sin}\epsilon)^2$ & $\frac{1}{9}$ & $\frac{8}{9}$ & $\frac{2}{9}$ & $\frac{1}{9}(\sqrt{3}\textrm{cos}\epsilon + \textrm{sin}\epsilon)^2$ \\
$\Xi^0$ & $\frac{1}{9}(\textrm{cos}\epsilon + \sqrt{3}\textrm{sin}\epsilon)^2$ & $\frac{2}{9}$ & $\frac{1}{9}$ & $\frac{8}{9}$ & $\frac{1}{9}(-\sqrt{3}\textrm{cos}\epsilon +\textrm{sin}\epsilon)^2$ \\
$\Xi^-$ & $\frac{1}{9}(-\textrm{cos}\epsilon + \sqrt{3}\textrm{sin}\epsilon)^2$& $\frac{2}{9}$ & $\frac{8}{9}$ & $\frac{1}{9}$ & $\frac{1}{9}(\sqrt{3}\textrm{cos}\epsilon + \textrm{sin}\epsilon)^2$ \\ \hline

\end{tabular}
\end{center}
\caption{Chiral SU(3) coefficients for the coupling of the octet baryons to decuplet ($T$) baryons through the pseudoscalar octet meson $\phi$.}
\label{tab:deccoup}
\end{table}

\begin{table}[htbf]
\begin{center}
\begin{tabular}{c c c c}
\hline
& \multicolumn{3}{c}{$\chi_{B\phi}$} \\
 & \multicolumn{3}{c}{$\pi^0$} \\ \hline
$p$ & \multicolumn{3}{c}{$\frac{1}{6}(2(D^2+3F^2)+(D^2+6DF-3F^2)\textrm{cos}(2\epsilon)-\sqrt{3}(D-3F)(D+F)\textrm{sin}(2\epsilon))$} \\
$n$ & \multicolumn{3}{c}{$\frac{1}{6}(2(D^2+3F^2)+(D^2+6DF-3F^2)\textrm{cos}(2\epsilon)+\sqrt{3}(D-3F)(D+F)\textrm{sin}(2\epsilon))$}  \\
$\Sigma^+$ & \multicolumn{3}{c}{$F^2 + F^2\textrm{cos}(2\epsilon)+\frac{2}{3}D\textrm{sin}\epsilon(2\sqrt{3} F\textrm{cos}\epsilon+ D \textrm{sin}\epsilon)$}  \\
$\Sigma^-$ & \multicolumn{3}{c}{$F^2 + F^2\textrm{cos}(2\epsilon)+\frac{2}{3}D\textrm{sin}\epsilon(-2\sqrt{3} F\textrm{cos}\epsilon+ D \textrm{sin}\epsilon)$}  \\
$\Xi^0$ & \multicolumn{3}{c}{$\frac{1}{6}(2(D^2+3F^2)+(D^2-6DF-3F^2)\textrm{cos}(2\epsilon)+\sqrt{3}(D+3F)(D-F)\textrm{sin}(2\epsilon))$} \\
$\Xi^-$ & \multicolumn{3}{c}{$\frac{1}{6}(2(D^2+3F^2)+(D^2-6DF-3F^2)\textrm{cos}(2\epsilon)-\sqrt{3}(D+3F)(D-F)\textrm{sin}(2\epsilon))$} \\ \hline \hline
 & \multicolumn{3}{c}{$\eta$} \\ \hline 
$p$ & \multicolumn{3}{c}{$\frac{1}{6}(2(D^2+3F^2)-(D^2+6DF-3F^2)\textrm{cos}(2\epsilon)+\sqrt{3}(D-3F)(D+F)\textrm{sin}(2\epsilon))$} \\
$n$ & \multicolumn{3}{c}{$\frac{1}{6}(2(D^2+3F^2)-(D^2+6DF-3F^2)\textrm{cos}(2\epsilon)-\sqrt{3}(D-3F)(D+F)\textrm{sin}(2\epsilon))$} \\
$\Sigma^+$ & \multicolumn{3}{c}{$\frac{2}{3}(D^2\textrm{cos}^2\epsilon-2\sqrt{3}DF\textrm{cos}\epsilon\textrm{sin}\epsilon+3F^2\textrm{sin}^2\epsilon)$} \\
$\Sigma^-$ & \multicolumn{3}{c}{$\frac{2}{3}(D^2\textrm{cos}^2\epsilon+2\sqrt{3}DF\textrm{cos}\epsilon\textrm{sin}\epsilon+3F^2\textrm{sin}^2\epsilon)$} \\
$\Xi^0$ & \multicolumn{3}{c}{$\frac{1}{6}(2(D^2+3F^2)+(-D^2+6DF+3F^2)\textrm{cos}(2\epsilon)-\sqrt{3}(D+3F)(D-F)\textrm{sin}(2\epsilon))$} \\
$\Xi^-$ & \multicolumn{3}{c}{$\frac{1}{6}(2(D^2+3F^2)+(-D^2+6DF+3F^2)\textrm{cos}(2\epsilon)+\sqrt{3}(D+3F)(D-F)\textrm{sin}(2\epsilon))$} \\ \hline \hline
\end{tabular}
\begin{tabular}{c c c c}
 & $\pi^\pm$ & $K^0$ & $K^\pm$  \\ \hline 
$p$ & $(D+F)^2$ & $(D-F)^2$ & $\frac{2}{3}(D^2+3F^2)$  \\
$n$ & $(D+F)^2$ & $\frac{2}{3}(D^2+3F^2)$ & $(D-F)^2$  \\
$\Sigma^+$ & $\frac{2}{3}(D^2+3F^2)$ & $(D-F)^2$ & $(D+F)^2$  \\
$\Sigma^-$ & $\frac{2}{3}(D^2+3F^2)$ & $(D+F)^2$ & $(D-F)^2$  \\
$\Xi^0$ & $(D-F)^2$ & $\frac{2}{3}(D^2+3F^2)$ & $(D+F)^2$  \\
$\Xi^-$ & $(D-F)^2$ & $(D+F)^2$ & $\frac{2}{3}(D^2+3F^2)$  \\ \hline
\end{tabular}
\end{center}
\caption{Chiral SU(3) coefficients for the coupling of the octet baryons to octet ($B$) baryons through the pseudoscalar octet meson $\phi$.}
\label{tab:octcoup}
\end{table}

It is now straightforward to write expressions for the baryon mass splittings as a function of quark mass only. All other free parameters, namely the SU(3) breaking parameters $\alpha$, $\beta$ and $\sigma$, as well as the regulator mass $\Lambda$, are specified by the isospin-averaged fit described previously.

To evaluate the baryon mass splittings at the physical point, we input the physical light-quark mass ratio $R:=\frac{m_u}{m_d}$.
The Gell-Mann-Oakes Renner relation suggests the definition
\begin{equation}
\omega = \frac{B(m_d-m_u)}{2} := \frac{1}{2} \frac{(1-R)}{(1+R)} m_{\pi_\textrm{\tiny{(phys)}}}^2,
\end{equation}
which allows us to define
\begin{subequations}
\begin{align}
Bm_u & =m_{\pi_\textrm{\tiny{(phys)}}}^2/2 - \omega, \\
Bm_d & =m_{\pi_\textrm{\tiny{(phys)}}}^2/2 + \omega, \\
Bm_s & =m_{K_\textrm{\tiny{(phys)}}}^2-m_{\pi_\textrm{\tiny{(phys)}}}^2/2.
\end{align}
\end{subequations}
%
%
%
%
Here, we take $m_{\pi_\textrm{\tiny{(phys)}}}=137.3$~MeV and $m_{K_\textrm{\tiny{(phys)}}}=497.5$~MeV to be the physical isospin-averaged meson masses~\cite{pdg}.

Evaluating the mass splitting expressions at these `physical' quark masses, with loop meson masses calculated using Equation~\ref{mesonmasses}, then gives our estimate of the baryon mass differences at the physical point. A discussion of the error analysis is given in section~\ref{subsec:syst}.

\section{Results}

The calculation outlined in the previous section, where the central value is obtained using the stated phenomenological estimates for $f$, $F$, $\mathcal{C}$ and $\delta$, gives
\begin{subequations}
\begin{align}
M_n-M_p & = (\omega / m_{\pi_\textrm{\tiny{(phys)}}}^2)(20.3 \pm 1.2) \textrm{ MeV}, \\
M_{\Sigma^-}- M_{\Sigma^+} & =  (\omega / m_{\pi_\textrm{\tiny{(phys)}}}^2)(52.6 \pm 2.0) \textrm{ MeV},\\
M_{\Xi^-}-M_{\Xi^0} & =  (\omega / m_{\pi_\textrm{\tiny{(phys)}}}^2)(32.3 \pm 1.6) \textrm{ MeV}.
\end{align}
\end{subequations}
The quoted uncertainties contain all statistical and systematic errors, discussed in the following section, combined in quadrature.

As input we take two recent estimates for the physical up-down quark mass ratio~\cite{Leutwyler1996,FLAG},
\begin{equation}
R:=\frac{m_u}{m_d} = 0.553 \pm 0.043, \textrm{~~and~~~} 0.47 \pm 0.04.
\label{eq:R}
\end{equation}
The first of these is determined by a fit to meson decay rates. We note that this value is compatible with more recent estimates of the ratio from $2+1$ and 3 flavor QCD and QED~\cite{Aubin2004,Blum2010}. The second is the result from the FLAG. For the two estimates for $R$, we find, respectively
\begin{subequations}
\begin{align}
M_n-M_p & = 2.9 \pm 0.4 \textrm{ MeV}, \textrm{~~and~~~} 3.7 \pm 0.4 \textrm{ MeV},\\
M_{\Sigma^-}- M_{\Sigma^+} & = 7.5 \pm 1.0 \textrm{ MeV},\textrm{~~and~~~} 9.5 \pm 0.9 \textrm{ MeV},\\
M_{\Xi^-}-M_{\Xi^0} & = 4.6 \pm 0.6 \textrm{ MeV},\textrm{~~and~~~} 5.8 \pm 0.6 \textrm{ MeV},
\end{align}
\end{subequations}
where uncertainties have been added in quadrature.

\subsection{Statistical and systematic uncertainties}
\label{subsec:syst}

The errors quoted are the result of a complete error analysis, taking into account the correlated uncertainties arising from all of the fit parameters, as well as propagating the quoted uncertainty in $R$. We estimate the systematic error in our result by considering variations of the regulator and allowing for deviation of the phenomenologically set parameters $f$, $F$, $\mathcal{C}$ and $\delta$ from their central values by $\pm 10 \%$.

Monopole, dipole, Gaussian and sharp cutoff regulators $u(k)$ are considered in our analysis. The variation of our final results as $u(k)$ is changed is of order 1$\%$ of our determined mass differences, and is included in the quoted error. The deviation as the parameters $f$, $F$, $\mathcal{C}$ and $\delta$ are perturbed is similarly small, and the statistical uncertainty arising from the fit to lattice data is smaller still.

In fact, the dominant uncertainty by an order of magnitude is that that arising from the quoted error band on $R$, the light quark mass ratio. It is clear that better estimates of this value will allow our results to be greatly improved in precision, without the need for further lattice data. Conversely, a precise determination of the electromagnetic contribution to the $n-p$ mass difference could possibly facilitate an improved estimate of $R$ by this method.

\section{Application to QCDSF-UKQCD lattice results}

The method described above can be applied equally to other sets of lattice data for the octet baryon masses. In particular, we consider recent $2+1$-flavor QCDSF-UKQCD lattice simulations~\cite{Bietenholz2011}, which follow a significantly different trajectory in the light-strange quark mass plane to the PACS-CS simulations. Instead of holding the strange quark mass fixed along the simulation trajectory, the QCDSF-UKQCD Collaboration holds the singlet quark mass ($m_K^2+m_\pi^2/2$) fixed.

\begin{figure}[hftb]
\begin{minipage}[t]{\linewidth}
\centering
\includegraphics[width=0.7\textwidth]{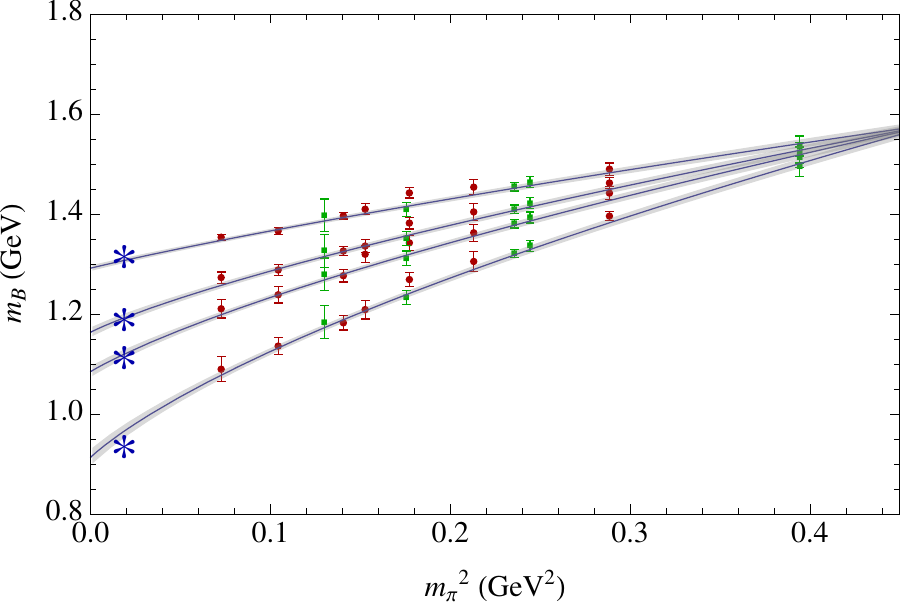}
\caption{Fit to the QCDSF-UKQCD baryon octet data. Error bands shown are purely statistical, and incorporate correlated uncertainties between all fit parameters. Note that the data shown has been corrected for finite volume and the simulation strange quark mass, which was somewhat different from the physical value at each point. The red circles and green squares lie on the singlet trajectory and the SU(3) symmetric line respectively, and the blue stars denote the physical points.}
\label{fig:octUKQCD}
\vspace{1cm}
\end{minipage}
\begin{minipage}[b]{0.45\linewidth}
\centering
\includegraphics[width=\textwidth]{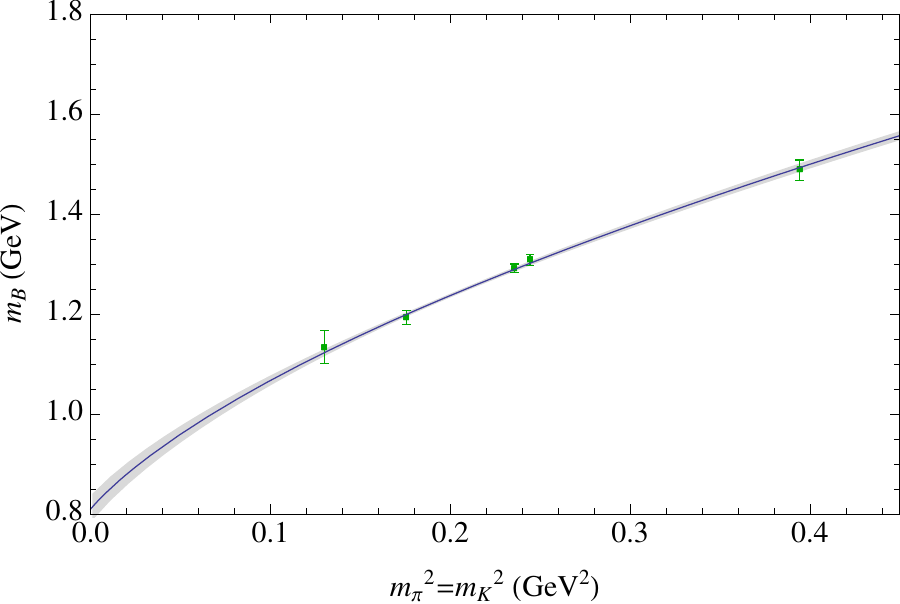}
\caption{Fit to the QCDSF-UKQCD baryon octet data, shown along the SU(3) symmetric line. Error bands are as in Figure~\ref{fig:octUKQCD}.}
\label{fig:su3line}
\end{minipage}
\hspace{0.5cm}
\begin{minipage}[b]{0.45\linewidth}
\centering
\includegraphics[width=\textwidth]{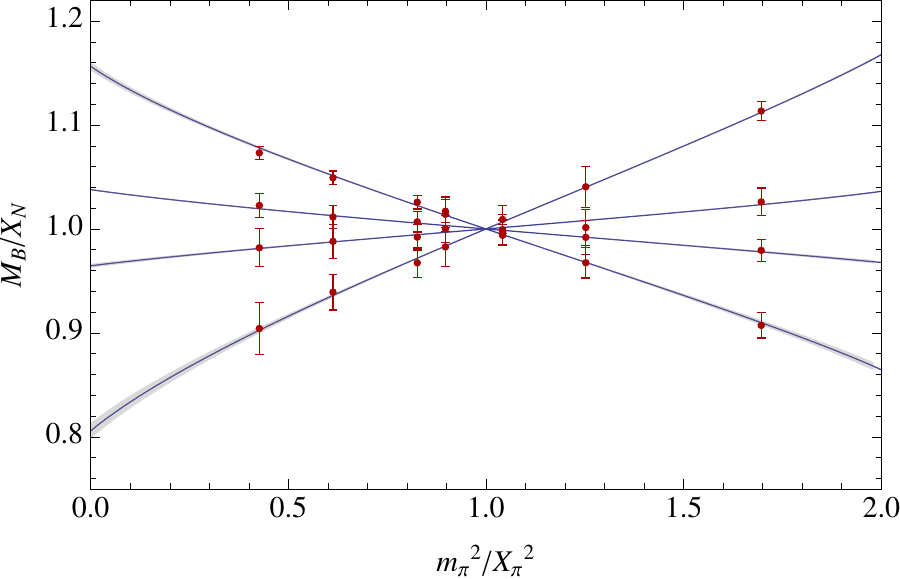}
\caption{Fit to the QCDSF-UKQCD baryon octet data, shown along the singlet trajectory. Error bands are as in Figure~\ref{fig:octUKQCD}.}
\label{fig:fan}
\end{minipage}
\end{figure}

We use simulation results from this collaboration which lie both along the `singlet' line and along the SU(3) symmetric line~\cite{Bietenholz2011}. Precisely as was done in our analysis of the PACS-CS lattice data, we calculate small finite-volume corrections. The lattice spacing $a=0.072(1)$~fm is determined by fixing $X_N=(1/3)(M_N+M_\Sigma+M_\Xi)$ to the experimental value at physical quark masses. This $a$ is somewhat lower than that quoted by the QCDSF-UKQCD Collaboration as we account for chiral curvature.
The quality of our fit to the isospin-averaged results is clearly excellent, with a $\chi^2/$dof of 0.6 and regulator mass $\Lambda=1.0\pm0.1$~GeV. The fit is shown in Figures~\ref{fig:octUKQCD}, \ref{fig:su3line} and \ref{fig:fan}, and the values of the octet baryon masses extrapolated to the physical point are given in Table~\ref{tab:barmassesUKQCD}. These are largely consistent with the physical values.

\begin{table}[hftb]
\begin{center}
\begin{tabular}{c r@{.}l c}
\toprule
$B$ & \multicolumn{2}{c}{Mass (GeV)} & Experimental \\ 
\midrule
$N$ & 0 &966(16)(10) & 0.939  \\
$\Lambda$ & 1 &112(13)(5) & 1.116  \\
$\Sigma$ & 1 &193(12)(4) & 1.193  \\
$\Xi$ & 1 &307(11)(0) & 1.318 \\
\bottomrule
\end{tabular}
\caption{Octet baryon masses based on a chiral extrapolation of the QCDSF-UKQCD data set. The first uncertainty quoted is statistical and the second results from the variation of various chiral parameters and the form of the UV regulator as described in the text. The experimental masses are shown for comparison.}
\label{tab:barmassesUKQCD}
\end{center}
\end{table}

Applying the method described in previous sections to the QCDSF-UKQCD lattice data gives results for the strong contribution to the octet baryon mass splittings which are consistent with our fit to the PACS-CS Collaboration simulation results. We find
\begin{subequations}
\begin{align}
M_n-M_p & = (\omega / m_{\pi_\textrm{\tiny{(phys)}}}^2)(16.6 \pm 1.2) \textrm{ MeV}, \\
M_{\Sigma^-}- M_{\Sigma^+} & =  (\omega / m_{\pi_\textrm{\tiny{(phys)}}}^2)(48.9 \pm 1.7) \textrm{ MeV},\\
M_{\Xi^-}-M_{\Xi^0} & =  (\omega / m_{\pi_\textrm{\tiny{(phys)}}}^2)(32.2 \pm 1.6) \textrm{ MeV},
\end{align}
\end{subequations}
corresponding to
\begin{subequations}
\label{eq:UKQCD}
\begin{align}
M_n-M_p & = 2.4 \pm 0.3 \textrm{ MeV}, \textrm{~~and~~~} 3.0 \pm 0.3 \textrm{ MeV},\\
M_{\Sigma^-}- M_{\Sigma^+} & = 7.0 \pm 0.9 \textrm{ MeV}, \textrm{~~and~~~} 8.9 \pm 0.8 \textrm{ MeV},\\
M_{\Xi^-}-M_{\Xi^0} & = 4.6 \pm 0.6 \textrm{ MeV}, \textrm{~~and~~~} 5.8 \pm 0.6 \textrm{ MeV},
\end{align}
\end{subequations}
when $R$ is set to each of the values given in Equation~\ref{eq:R}. 

The QCDSF-UKQCD Collaboration has also recently presented a determination of the strong contribution to the baryonic mass splittings, based on these lattice simulations~\cite{Horsley2012}. That calculation used a linear and quadratic SU(3) flavor symmetry breaking expansion in the quark masses. As the meson and baryon octet expansion coefficients depend only on the average quark mass, provided these are kept constant this method allows for an estimation of the baryon mass splittings at the physical point using the available $2+1$-flavor lattice results.
The results reported in that work are:
\begin{subequations}
\label{eq:theirs}
\begin{align}
M_n-M_p & = 3.13 \pm 0.15 \pm 0.53 \textrm{ MeV}, \\
M_{\Sigma^-}- M_{\Sigma^+} & = 8.10 \pm 0.14 \pm 1.35 \textrm{ MeV},\\
M_{\Xi^-}-M_{\Xi^0} & = 4.98 \pm 0.10 \pm 0.84 \textrm{ MeV}.
\end{align}
\end{subequations}
The first uncertainty quoted in Equation~\ref{eq:theirs} is statistical, while the second allows for violations of Dashen's theorem.

\section{Discussion}

We have carried out an analysis of isospin-averaged lattice simulations for octet baryon masses using a formal chiral expansion based on broken SU(3) symmetry. Using the resulting expansion coefficients one can evaluate the strong contribution to the baryon mass splittings.
Our results, based on analyses of PACS-CS and QCDSF-UKQCD lattice data sets, are summarized in Table~\ref{tab:summary}. Both calculations yield compatible values, despite significant differences between the two lattice simulations, including in particular different lattice sizes, lattice spacings, and different methods of determining these spacings. Of course, as emphasized previously, the two simulations also follow quite different trajectories in $m_\pi-m_K$ space.

Furthermore, we note that the results of the QCDSF-UKQCD Collaboration analysis of their data are are entirely consistent with our values. While the approach taken by the QCDSF-UKQCD group makes use of only that lattice data calculated along a trajectory which holds the average quark mass constant, we have also included in our fit the data from that collaboration which lies away from this line. This contributes to our reduced uncertainties. We also point out that both methods require some theoretical input: we input the up-down quark mass ratio $R$, while the the Horsely \textit{et al.} calculation uses Dashen's theorem (with some uncertainty) to estimate `pure QCD' meson masses at the physical point. The clear consistency between the two independent calculations is encouraging.

\begin{table}[hftb]
\begin{center}
\begin{tabular}{l c c c}
\toprule
\multicolumn{1}{c}{$\Delta_{m_d-m_u}$~(MeV)} & $M_n-M_p$ & $M_{\Sigma^-}-M_{\Sigma^+}$ & $M_{\Xi^-}-M_{\Xi^0}$ \\ 
\midrule
Chiral (PACS-CS) & 2.9(4) & 7.5(10) & 4.6(6)   \\
Chiral (QCDSF-UKQCD) & 2.4(3) & 7.0(9) & 4.6(6)   \\
QCDSF-UKQCD & 3.13(55) & 8.10(136) & 4.98(86)   \\
Exp. \& EM (traditional) & 2.05(30) & 7.91(30) & 5.99(30)   \\
Exp. \& EM (WLCM) & 2.60(47) &  & \\
\bottomrule
\end{tabular}
\caption{Up-down quark mass contribution to octet baryon mass splittings. Lines 1 and 2 show the results of our chiral extrapolation of PACS-CS and QCDSF-UKQCD lattice data respectively, with the up-down quark mass ratio set to $R:=m_u/m_d=0.553(43)$. Line 3 shows the QCDSF-UKQCD Collaboration analysis of their data as described in the previous section, while lines 4 and 5 give estimates deduced from the total mass splittings and electromagnetic contributions, as determined by Gasser and Leutwyler (traditional) or Walker-Loud~\textit{et al.} (WLCM).}
\label{tab:summary}
\end{center}
\end{table}

While our results and those of the QCDSF-UKQCD Collaboration are consistent with both the traditional and Walker-Loud~\textit{et al.} (WLCM) determinations of the strong contribution from the electromagnetic component (see Table~\ref{tab:summary}), it is clear that determinations of all octet baryon electromagnetic mass splittings using the WLCM analysis would be of considerable interest. We also emphasize that while more lattice data for isospin-averaged octet baryon masses, on larger lattice volumes, would allow the uncertainties of our calculation to be somewhat reduced, the dominant uncertainty in our calculation arises from the up-down quark mass ratio. A more precise value of $m_u/m_d$ could reduce the uncertainty of our determination by as much as an order of magnitude.
Conversely, direct lattice determinations of the electromagnetic contributions to the mass splittings, with the analysis presented here, may act to significantly improve the value of $m_u/m_d$.

\section*{Acknowledgements}

This work was supported by the University of Adelaide and the Australian Research Council through the ARC Centre of Excellence for Particle Physics at the Terascale and grants FL0992247 (AWT) and DP110101265 (RDY).

\section*{References}

\bibliographystyle{model1-num-names}
\bibliography{MassSplitBib}

\begin{thebibliography}{28}
\expandafter\ifx\csname natexlab\endcsname\relax\def\natexlab#1{#1}\fi
\providecommand{\bibinfo}[2]{#2}
\ifx\xfnm\relax \def\xfnm[#1]{\unskip,\space#1}\fi
\bibitem[{Mohr et~al.(2012)Mohr, Taylor, and Newell}]{Mohr2012}
\bibinfo{author}{P.~J. Mohr}, \bibinfo{author}{B.~N. Taylor},
  \bibinfo{author}{D.~B. Newell},
\newblock \bibinfo{title}{{CODATA Recommended Values of the Fundamental
  Physical Constants: 2010}}  (\bibinfo{year}{2012})
  \bibinfo{pages}{[arXiv:1203.5425]}.
\bibitem[{Beringer et~al.(2012)}]{pdg}
\bibinfo{author}{J.~Beringer}, et~al.,
\newblock \bibinfo{title}{{[Particle Data Group]}},
\newblock \bibinfo{journal}{Phys. Rev. D} \bibinfo{volume}{86}
  (\bibinfo{year}{2012}) \bibinfo{pages}{010001}.
\bibitem[{Beane et~al.(2007)Beane, Orginos, and Savage}]{Beane2006}
\bibinfo{author}{S.~R. Beane}, \bibinfo{author}{K.~Orginos},
  \bibinfo{author}{M.~J. Savage},
\newblock \bibinfo{title}{{Strong-isospin violation in the neutron-proton mass
  difference from fully-dynamical lattice QCD and PQQCD}},
\newblock \bibinfo{journal}{Nucl. Phys. B} \bibinfo{volume}{768}
  (\bibinfo{year}{2007}) \bibinfo{pages}{38--50}.
\bibitem[{Blum et~al.(2010)}]{Blum2010}
\bibinfo{author}{T.~Blum}, et~al.,
\newblock \bibinfo{title}{{Electromagnetic mass splittings of the low lying
  hadrons and quark masses from $2+1$ flavor lattice QCD$+$QED}},
\newblock \bibinfo{journal}{Phys. Rev. D} \bibinfo{volume}{82}
  (\bibinfo{year}{2010}) \bibinfo{pages}{094508}.
\bibitem[{Walker-Loud(2010)}]{WalkerLoud2010}
\bibinfo{author}{A.~Walker-Loud},
\newblock \bibinfo{title}{{Towards a direct lattice calculation of $m_d -
  m_u$}},
\newblock \bibinfo{journal}{PoS} \bibinfo{volume}{243} (\bibinfo{year}{2010}).
\bibitem[{de~Divitiis et~al.(2011)}]{Divitiis2011}
\bibinfo{author}{G.~M. de~Divitiis}, et~al.,
\newblock \bibinfo{title}{{Isospin breaking effects due to the up-down mass
  difference in Lattice QCD}}  (\bibinfo{year}{2011})
  \bibinfo{pages}{[arXiv:1110.6294]}.
\bibitem[{Horsley et~al.(2012)}]{Horsley2012}
\bibinfo{author}{R.~Horsley}, et~al.,
\newblock \bibinfo{title}{{Isospin breaking in octet baryon mass splittings}}
  (\bibinfo{year}{2012}) \bibinfo{pages}{[arXiv:1206.3156]}.
\bibitem[{Duncan et~al.(1996)Duncan, Eichten, and Thacker}]{Duncan1996}
\bibinfo{author}{A.~Duncan}, \bibinfo{author}{E.~Eichten},
  \bibinfo{author}{H.~Thacker},
\newblock \bibinfo{title}{{Electromagnetic Splittings and Light Quark Masses in
  Lattice QCD}},
\newblock \bibinfo{journal}{Phys. Rev. Lett.} \bibinfo{volume}{76}
  (\bibinfo{year}{1996}) \bibinfo{pages}{3894--3897}.
\bibitem[{Basak et~al.(2008)}]{Basak2008}
\bibinfo{author}{S.~Basak}, et~al.,
\newblock \bibinfo{title}{{Electromagnetic splittings of hadrons from improved
  staggered quarks in full QCD}},
\newblock \bibinfo{journal}{PoS LATTICE} \bibinfo{volume}{127}
  (\bibinfo{year}{2008}).
\bibitem[{Portelli(2011)}]{Portelli2011}
\bibinfo{author}{A.~Portelli},
\newblock \bibinfo{title}{{Systematic errors in partially-quenched QCD plus QED
  lattice simulations}},
\newblock \bibinfo{journal}{PoS LATTICE} \bibinfo{volume}{136}
  (\bibinfo{year}{2011}).
\bibitem[{Glaessle and Bali(2011)}]{Glaessle2011}
\bibinfo{author}{B.~Glaessle}, \bibinfo{author}{G.~S. Bali},
\newblock \bibinfo{title}{{Electromagnetic corrections to pseudoscalar decay
  constants}},
\newblock \bibinfo{journal}{PoS LATTICE} \bibinfo{volume}{282}
  (\bibinfo{year}{2011}).
\bibitem[{Aoki et~al.(2009)}]{Aoki2009}
\bibinfo{author}{S.~Aoki}, et~al.,
\newblock \bibinfo{title}{{2+1 Flavor Lattice QCD toward the Physical Point}},
\newblock \bibinfo{journal}{Phys. Rev. D} \bibinfo{volume}{79}
  (\bibinfo{year}{2009}) \bibinfo{pages}{034503}.
\bibitem[{Bietenholz et~al.(2011)}]{Bietenholz2011}
\bibinfo{author}{W.~Bietenholz}, et~al.,
\newblock \bibinfo{title}{{Flavor blindness and patterns of flavor symmetry
  breaking in lattice simulations of up, down, and strange quarks}},
\newblock \bibinfo{journal}{Phys. Rev. D} \bibinfo{volume}{84}
  (\bibinfo{year}{2011}) \bibinfo{pages}{054509}.
\bibitem[{Gasser and Leutwyler(1982)}]{Gasser1982}
\bibinfo{author}{J.~Gasser}, \bibinfo{author}{H.~Leutwyler},
\newblock \bibinfo{title}{{Quark Masses}},
\newblock \bibinfo{journal}{Phys. Rept.} \bibinfo{volume}{87}
  (\bibinfo{year}{1982}) \bibinfo{pages}{77--169}.
\bibitem[{Walker-Loud et~al.(2012)Walker-Loud, Carlson, and
  Miller}]{Walker-Loud2012}
\bibinfo{author}{A.~Walker-Loud}, \bibinfo{author}{C.~E. Carlson},
  \bibinfo{author}{G.~A. Miller},
\newblock \bibinfo{title}{{The Electromagnetic Self-Energy Contribution to $M_p
  - M_n$ and the Isovector Nucleon Magnetic Polarizability}}
  (\bibinfo{year}{2012}) \bibinfo{pages}{[arXiv:1203.0254]}.
\bibitem[{Young and Thomas(2010)}]{Young2010}
\bibinfo{author}{R.~D. Young}, \bibinfo{author}{A.~W. Thomas},
\newblock \bibinfo{title}{Octet baryon masses and sigma terms from an su(3)
  chiral extrapolation},
\newblock \bibinfo{journal}{Phys. Rev. D} \bibinfo{volume}{81}
  (\bibinfo{year}{2010}) \bibinfo{pages}{014503}.
\bibitem[{Shanahan et~al.(2011)Shanahan, Thomas, and Young}]{Shanahan2011}
\bibinfo{author}{P.~E. Shanahan}, \bibinfo{author}{A.~W. Thomas},
  \bibinfo{author}{R.~D. Young},
\newblock \bibinfo{title}{{Mass of the H-dibaryon}},
\newblock \bibinfo{journal}{Phys. Rev. Lett.} \bibinfo{volume}{107}
  (\bibinfo{year}{2011}) \bibinfo{pages}{092004}.
\bibitem[{Shanahan et~al.(2012)Shanahan, Thomas, and Young}]{Shanahan2012}
\bibinfo{author}{P.~E. Shanahan}, \bibinfo{author}{A.~W. Thomas},
  \bibinfo{author}{R.~D. Young},
\newblock \bibinfo{title}{{Sigma terms from an SU(3) chiral extrapolation}}
  (\bibinfo{year}{2012}) \bibinfo{pages}{[arXiv:1205.5365]}.
\bibitem[{Stuckey and Birse(1997)}]{Stuckey1997}
\bibinfo{author}{R.~E. Stuckey}, \bibinfo{author}{M.~C. Birse},
\newblock \bibinfo{title}{{Baryon masses in a chiral expansion with meson -
  baryon form factors}},
\newblock \bibinfo{journal}{J. Phys. G} \bibinfo{volume}{23}
  (\bibinfo{year}{1997}) \bibinfo{pages}{29}.
\bibitem[{Donoghue et~al.(1999)}]{Donoghue1999}
\bibinfo{author}{J.~F. Donoghue}, et~al.,
\newblock \bibinfo{title}{{SU(3) baryon chiral perturbation theory and long
  distance regularization}},
\newblock \bibinfo{journal}{Phys. Rev. D} \bibinfo{volume}{59}
  (\bibinfo{year}{1999}) \bibinfo{pages}{036002}.
\bibitem[{Young et~al.(2002)Young, Leinweber, Thomas, and Wright}]{Young2002}
\bibinfo{author}{R.~D. Young}, \bibinfo{author}{D.~B. Leinweber},
  \bibinfo{author}{A.~W. Thomas}, \bibinfo{author}{S.~V. Wright},
\newblock \bibinfo{title}{{Chiral Analysis of Quenched Baryon Masses}},
\newblock \bibinfo{journal}{Phys. Rev. D} \bibinfo{volume}{66}
  (\bibinfo{year}{2002}) \bibinfo{pages}{094507}.
\bibitem[{Young et~al.(2003)Young, Leinweber, and Thomas}]{YOung2003}
\bibinfo{author}{R.~D. Young}, \bibinfo{author}{D.~B. Leinweber},
  \bibinfo{author}{A.~W. Thomas},
\newblock \bibinfo{title}{{Convergence of Chiral Effective Field Theory}},
\newblock \bibinfo{journal}{Prog. Part. Nucl. Phys.} \bibinfo{volume}{50}
  (\bibinfo{year}{2003}) \bibinfo{pages}{399--417}.
\bibitem[{Leinweber et~al.(2004)Leinweber, Thomas, and Young}]{Leinweber2004}
\bibinfo{author}{D.~B. Leinweber}, \bibinfo{author}{A.~W. Thomas},
  \bibinfo{author}{R.~D. Young},
\newblock \bibinfo{title}{{Physical Nucleon Properties from Lattice QCD}},
\newblock \bibinfo{journal}{Phys. Rev. Lett.} \bibinfo{volume}{92}
  (\bibinfo{year}{2004}) \bibinfo{pages}{242002}.
\bibitem[{Walker-Loud(2005)}]{WalkerLoud2004}
\bibinfo{author}{A.~Walker-Loud},
\newblock \bibinfo{title}{{Octet baryon masses in partially quenched chiral
  perturbation theory}},
\newblock \bibinfo{journal}{Nucl. Phys. A} \bibinfo{volume}{747}
  (\bibinfo{year}{2005}) \bibinfo{pages}{476--507}.
\bibitem[{Amoros et~al.(2001)Amoros, Bijnens, and Talavera}]{Amoros2001}
\bibinfo{author}{G.~Amoros}, \bibinfo{author}{J.~Bijnens},
  \bibinfo{author}{P.~Talavera},
\newblock \bibinfo{title}{{QCD isospin breaking in meson masses, decay
  constants and quark mass ratios}},
\newblock \bibinfo{journal}{Nucl. Phys. B} \bibinfo{volume}{602}
  (\bibinfo{year}{2001}) \bibinfo{pages}{87--108}.
\bibitem[{Leutwyler(1996)}]{Leutwyler1996}
\bibinfo{author}{H.~Leutwyler},
\newblock \bibinfo{title}{{The Ratios of the Light Quark Masses}},
\newblock \bibinfo{journal}{Phys. Lett. B} \bibinfo{volume}{378}
  (\bibinfo{year}{1996}) \bibinfo{pages}{313--318}.
\bibitem[{Colangelo et~al.(2011)}]{FLAG}
\bibinfo{author}{G.~Colangelo}, et~al.,
\newblock \bibinfo{title}{Review of lattice results concerning low-energy
  particle physics},
\newblock \bibinfo{journal}{Eur. Phys. J. C} \bibinfo{volume}{71}
  (\bibinfo{year}{2011}) \bibinfo{pages}{1--76}.
\bibitem[{Aubin et~al.(2004)}]{Aubin2004}
\bibinfo{author}{C.~Aubin}, et~al.,
\newblock \bibinfo{title}{{Light pseudoscalar decay constants, quark masses,
  and low energy constants from three-flavor lattice QCD}},
\newblock \bibinfo{journal}{Phys. Rev. D} \bibinfo{volume}{70}
  (\bibinfo{year}{2004}) \bibinfo{pages}{114501}.

\end{thebibliography}


\end{document}